\documentclass[aps,prl,twocolumn,preprintnumbers,superscriptaddress]{revtex4}
\usepackage{graphicx}
\usepackage{epstopdf}
\usepackage{amsmath}
\usepackage{amsfonts}
\usepackage{amssymb}
\usepackage{appendix}
\usepackage{enumerate}
\usepackage{comment}
\usepackage{bbold}
\usepackage[shortlabels]{enumitem}
\usepackage{color}
\usepackage{slashed}
\usepackage{subfigure}
\usepackage{setspace}
\usepackage{footnote}
\usepackage{multirow}
\usepackage[colorlinks = true,
            linkcolor = blue,
            urlcolor  = blue,
            citecolor = blue,
            anchorcolor = blue]{hyperref}
\usepackage[capitalize]{cleveref}
\usepackage{braket}
\usepackage[compat=1.1.0]{tikz-feynman}
\usepackage{multirow}
\usepackage{physics}
\usepackage{feynmp-auto}

\def\bar{\overline}

\begin{document}

\singlespacing

\preprint{FERMILAB-PUB-21-681-T}
\preprint{NUHEP-TH/21-18}
%\preprint{version:v4}

\title{Resonances in $\bar\nu_e-e^-$ scattering below a TeV}

\author{Vedran Brdar} 
\email{vedran.brdar@northwestern.edu}
\affiliation{Northwestern University, Department of Physics \& Astronomy, 2145 Sheridan Road, Evanston, IL 60208, USA}
\affiliation{Theoretical Physics Department, Fermilab, P.O. Box 500, Batavia, IL 60510, USA}
\author{Andr\'{e} de Gouv\^{e}a} 
\email{degouvea@northwestern.edu}
\affiliation{Northwestern University, Department of Physics \& Astronomy, 2145 Sheridan Road, Evanston, IL 60208, USA}
\author{Pedro A.~N.~Machado}
\email{pmachado@fnal.gov}
\affiliation{Theoretical Physics Department, Fermilab, P.O. Box 500, Batavia, IL 60510, USA}
\author{Ryan Plestid}
\email{rpl225@g.uky.edu }
\affiliation{Department of Physics and Astronomy, University of Kentucky, Lexington, KY 40506, USA}
\affiliation{Theoretical Physics Department, Fermilab, P.O. Box 500, Batavia, IL 60510, USA}

\begin{abstract}
%%%%%%%%%%%%%%
We consider the resonant production and detection of charged mesons in existing and near-future neutrino scattering experiments with $E_\nu \lesssim 1$ TeV, characteristic of high-energy atmospheric neutrinos or collider-sourced neutrino beams. The most promising candidate is the reaction $\bar{\nu}_e e^-\rightarrow \rho^-\rightarrow \pi^- \pi^0$. We discuss detection prospects at FASER$\nu$, the LHC's forward physics facility with nuclear emulsion (FASER$\nu$2) and liquid argon detectors (FLArE) and estimate the number of expected resonance-mediated events in the existing data set of IceCube. We also outline possible detection strategies for the different experimental environments. We predict dozens of events at the forward physics facility and identify cuts with order one signal efficiency that could potentially suppress backgrounds at FASER$\nu$, yielding a signal-to-background ratio larger than 1. Antineutrino-induced $s$-channel meson resonances are yet unobserved Standard Model scattering processes which offer a realistic target for near-term experiments. 
%%%%%%%%%%%%%%%%%
\end{abstract}

\maketitle

\textbf{Introduction.} Resonances are among the most distinctive and historically important observables in particle physics. In particular,  $s$-channel resonances imprint the unmistakable signature of on-shell particles onto experimental data. % observables such as production cross sections.
Famous examples include the discovery of the J$/\psi$ meson~\cite{SLAC-SP-017:1974ind, E598:1974sol}, the $Z$ boson \cite{UA1:1983mne,UA2:1983mlz,UA2:1983tsx} and the Higgs boson \cite{CMS:2012qbp,ATLAS:2012yve}. 

Relative to the rest of particle physics, resonances have played a minor role in the history of neutrino scattering physics.
Typical experimental configurations aim neutrino beams at detectors composed of electrons and nucleons. Since neutrinos carry lepton number, the only relevant Standard Model (SM) $s$-channel reaction is $\bar{\nu}_e e^- \rightarrow {R}^-$ where ${R}^-$  is a charged state with zero baryon and lepton number. 
The canonical example is the Glashow resonance $\bar{\nu}_e e^- \rightarrow W^- $ \cite{Glashow:1960zz}, which was recently identified as the likely origin of one candidate event in the IceCube neutrino observatory~\cite{IceCube:2021rpz}. 
The production of $W$ bosons from antineutrinos scattering on electrons at rest requires ultra-high energies, $E_\nu = M_W^2/2 m_e = 6.3$~PeV. As no terrestrial sources of neutrinos can reach such high energies, only the highest energy astrophysical neutrinos can produce a Glashow resonance. 

A natural question is whether neutrino beams can be used to produce observable resonances at lower energies. Of particular interest are neutrinos from the LHC and the associated forward physics facility (FPF) \cite{Anchordoqui:2021ghd}. For $E_\nu \lesssim 1$~TeV, $s = 2 m_e E_\nu <$~few GeV$^2$ charged resonances around or below the GeV scale, carrying neither baryon nor lepton number, are in principle accessible. These are the light (and, perhaps, charmed) charged mesons. The idea of detecting the $\rho^{-}$-meson resonance at IceCube was first sketched in \cite{Paschos:2002sj}. More recently, neutral meson resonances were considered in the context of the cosmic neutrino background~\cite{Dev:2021tlo}.  

%\begin{figure}
%\begin{fmffile}{simple_labels}
%\begin{fmfgraph*}(200,120)
%\fmfleft{i1,i2}
%\fmfright{o1,o2}
%\fmflabel{$e^-$}{i1}
%\fmflabel{$\bar{\nu}_e$}{i2}
%\fmflabel{$\pi^-$}{o1}
%\fmflabel{$\pi^0$}{o2}
%\fmf{dashes}{i1,v1,i2}
%\fmf{dashes}{o1,v2,o2}
%\fmf{dbl_wiggly,label=$\rho^-$}{v1,v2}
%\end{fmfgraph*}
%\end{fmffile}
%\vspace{12pt}
%\caption{Resonant production of a $\rho^-$ meson resulting in a $\pi^-\pi^0$ final state.}
%\label{fig:feynman}
%\end{figure}

%%%%%%%%%%%%%%%%%%%%%%%%%%%%%%%%%%%%%%%%%%%%%%%%%%%%%%%%%%%%%%%%%%%%%%%%%%%%%%%%%%%%%%%%%%%%%%%%%
 \begin{figure}[!]
	\centering
	\includegraphics[width=0.5\textwidth]{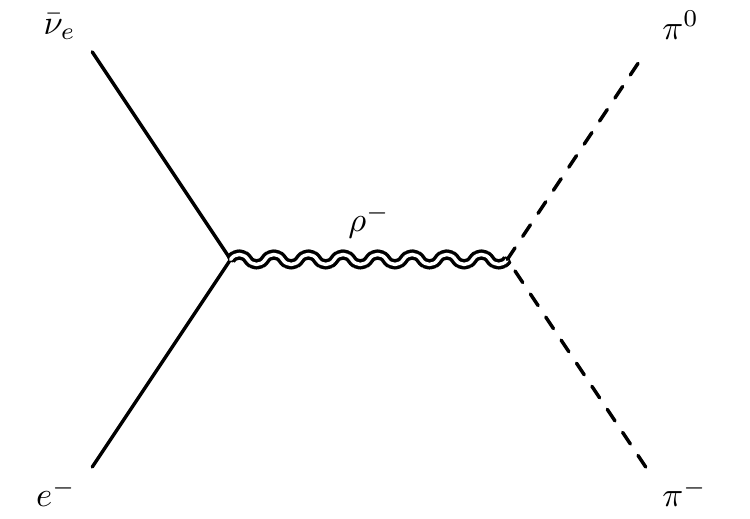} 
	\caption{Resonant production of a $\rho^-$ meson resulting in a $\pi^-\pi^0$ final state.}
	\label{fig:feynman}
\end{figure}
 %%%%%%%%%%%%%%%%%%%%%%%%%%%%%%%%%%%%%%%%%%%%%%%%%%%%%%%%%%%%%%%%%%%%%%%%%%%%%%%%%%%%%%%%%%%%%%%%
  
In this paper we consider the production of charged mesons $\mathfrak{m}$ via $\bar{\nu}_e e^-\rightarrow \mathfrak{m}$; these may be considered the low-energy analogs of the Glashow resonance, see \cref{fig:feynman}. From the perspective of neutrino physics, these events are interesting because they potentially provide event-by-event flavor tagging and because of their unique event topology. In particular,  $\bar{\nu}_e e^- \rightarrow \mathfrak{m}$ is expected to ``look'' very different from the more abundant deep inelastic scattering (DIS) process.
We estimate the number of resonant events in FASER$\nu$ \cite{FASER:2020gpr}, FLArE \cite{Batell:2021aja}, and IceCube \cite{IceCube:2016zyt}. We further discuss event topologies in Cherenkov, nuclear emulsion, and liquid argon time projection chamber (LArTPC) detectors, outlining distinctive signal characteristics, potential strategies for suppressing backgrounds, and comment on the prospect of measuring meson resonances in near-term neutrino experiments. 

\begin{figure}[!]
	\centering
	\includegraphics[width=0.5\textwidth]{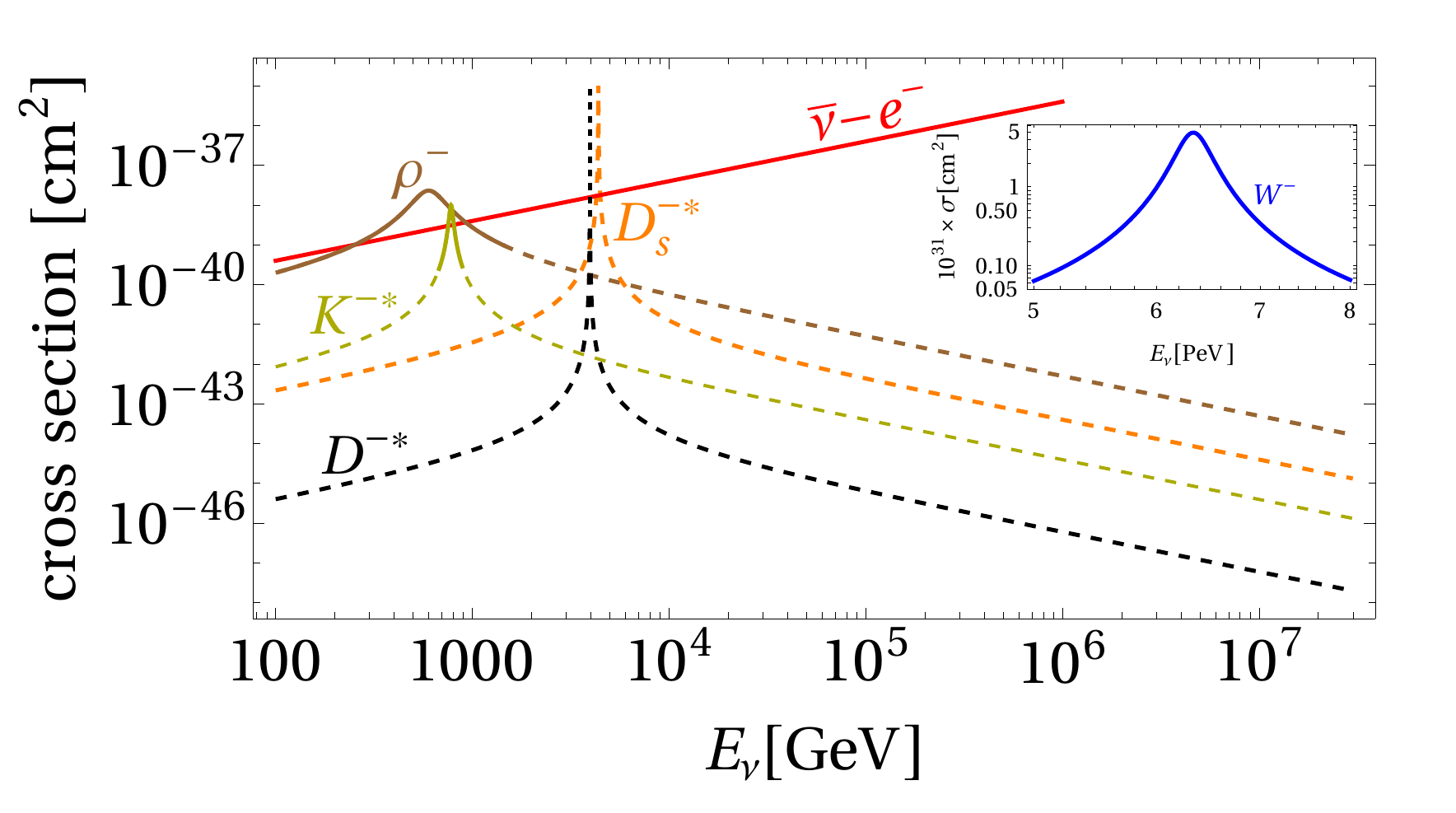} 
	\caption{Cross section for vector meson resonances, $\bar{\nu}_e e^-\to \mathfrak{m} \to$~hadrons. For comparison, the corresponding cross sections for $W$ boson production (inset) and $\bar{\nu}_e e^- \rightarrow \bar{\nu}_e e^-$ elastic  scattering (red line) are also included.}
	\label{fig:cross}
\end{figure}

\textbf{General Theory.}
Within the Breit-Wigner approximation the cross section for $\bar{\nu}_e e^-\to \mathfrak{m} \to X$ is given by \cite{IceCube:2021rpz,Barger:2014iua,Babu:2019vff}
\begin{align} \label{eq:BW}
%  &\sigma_{\rm res} =\frac{2J+1}{3} \frac{4\pi}{k^2}
 % \bigg[\frac{\Gamma^2/4}{(E-M)^2+\Gamma^2/4} \bigg]  Br_{\rm in} Br_{X}\,,
  & \sigma_{\rm res}= (2J+1) 8 \pi \Gamma^2 {\rm Br}_{\rm in} {\rm Br}_{\rm fi} \frac{s/M^2}{(s-M^2)^2+M^2\Gamma^2}\,,
\end{align}
where $J$ is the spin of the resonance, $\sqrt{s}$ is the center-of-mass energy, $M$ is the mass of the resonance, $\Gamma$ is the width of the resonance, ${\rm Br}_{\rm in}=\Gamma(\mathfrak{m}\to  \bar{\nu}_e e^-)/\Gamma$, and ${\rm Br}_{\rm fi}$ is the equivalent expression for the decay of a meson into final state $X$. 
The production of pseudoscalar-mesons, e.g.\ $\pi^-$ or $K^-$, is chirality suppressed by a factor $m_e^2/M^2$ and can therefore be safely neglected~\footnote{For reference, we expect fewer than $10^{-5}$ $\pi^-$ events at FLArE-100 and FASER$\nu$2.}. 
Consequently, the lightest meson with a sizeable production cross section is the charged vector $\rho^-$.  
Neglecting the electron mass,  the decay rate of vector mesons into antineutrinos and electrons is given by
\begin{align} \label{eq:decayvector}
  \Gamma(\mathfrak{m} \to \bar{\nu}_e e^-) =\frac{ G_F^2 }{12\pi}f^2 M^3
  |V_{\rm CKM}|^2\,,
\end{align}
where $V_{\rm CKM}$ is the relevant element of the quark mixing matrix, $G_F$ is the Fermi constant, and $f$ is the meson decay constant \cite{Donoghue:1992dd}. We use the values of $f$ listed in \cite{Chang:2018aut,Bondarenko:2018ptm}, e.g.\ $f_{\rho^-} = 0.21\, \text{GeV}$. %and stress that in \cref{eq:decayvector} the meson decay constants have dimensions of energy. 
\cref{fig:cross} depicts the cross sections for the production of different vector meson resonances computed using \cref{eq:BW}, where we set ${\rm Br}_{\rm fi}=1$, along with the corresponding cross section for on-shell $W^-$ boson production (in the inset) and low-energy $\bar{\nu}_e e^- \rightarrow \bar{\nu}_e e^-$ elastic  scattering.
%%%%%%%%%%%%%%%%%%%%%%%%%%%%%%%%%%%%%%%%%%%%%%%%%%%%%%%%%%%%%%%%%%%%%%%%%%%%%%%%%%%%%%%%%%%%%%%%%
 %%%%%%%%%%%%%%%%%%%%%%%%%%%%%%%%%%%%%%%%%%%%%%%%%%%%%%%%%%%%%%%%%%%%%%%%%%%%%%%%%%%%%%%%%%%%%%%%

We estimate the number of resonance-mediated events, $N_{\rm res}$, at a given experimental setup: 
\begin{align}
  N_{\rm res }=N_e
  \int_{\frac{(M-n\Gamma/2)^2}{(2m_e)}}^{\frac{(M+n\Gamma/2)^2}{(2m_e)}}
  \Phi(E_\nu) \sigma_{\rm res}(E_\nu) dE_\nu\,,
  \label{eq:rates}
\end{align}
where $N_e$ is the number of electrons in the fiducial volume of the detector, $\Phi$ is the antineutrino flux and $n$ denotes the number of charged meson widths $\Gamma$ across which the integral is evaluated.
For extremely narrow resonances, we checked that our results are equivalent to those obtained using the narrow-width approximation. For the $\rho^-$ resonance, it is well known that the near-peak structure of the resonance deviates substantially from a Breit-Wigner shape. The simplest way to circumvent this difficulty is to calculate $\bar{\nu}_e e^-\to \pi^0\pi^-$ in terms of form factors defined via $\mel{\pi^-(k_1)\pi^0(k_2)}{V_\mu}{0}= (k_1-k_2)_\mu F_1(q^2)+ (k_1+k_2)_\mu F_2(q^2)$, where $q=k_1-k_2$ \cite{Passera:2011ae,Czarnecki:2019iwz}. All terms that depend on $F_2$ are lepton-mass suppressed $O(m_e^2/q^2)$ and can be safely neglected. The form factor $F_1$ can be extracted from $e^+e^-\rightarrow \pi^+\pi^-$ data, but the extraction process requires isospin corrections and the subtraction of the $\omega(782)$ contribution. We instead use a simple isospin-symmetric chiral-perturbation-theory-inspired model with no adjustable parameters that agrees with $e^+e^-$ data at the $10\%$ level and explicitly excludes the $\omega(782)$ contribution \cite{Guerrero:1997ku}. 
Ultimately, this more detailed procedure yields event rates that are very similar to the ones obtained with \cref{eq:BW}, with differences typically below $10\%$. 
For obtaining the event numbers in \cref{table}, we employ this improved model for the cross section.

\textbf{The Weak R-ratio.} %Although somewhat tangential to our main line of discussion, 
The possibility that $\bar{\nu}_e e^- \to {\rm hadrons}$ may yield complementary information to $e^+e^- \to {\rm hadrons}$ is intriguing.  The $R$-ratio, defined as $R=\sigma(e^+e^- \to {\rm hadrons})\big/\sigma(e^+ e^- \to \mu^+\mu^-)$ (see \cite{ParticleDataGroup:2020ssz}, pages 699-700), is, for example, a key input for estimating the non-perturbative hadronic vacuum polarization \cite{Aoyama:2020ynm} and has received increased recent interest in light of the new precision measurements of the $(g-2)_\mu$ anomaly \cite{Muong-2:2006rrc,Muong-2:2021ojo}.  Given the prominent position occupied by the conventional $R$-ratio, we are motivated to define its weak interaction analog
\begin{equation}
    R_W = \frac{\sigma(\bar{\nu}_e e^- \to {\rm hadrons})}{\sigma(\bar{\nu}_e e^- \to \bar{\nu}_\mu \mu^-)}~.
\end{equation}

\begin{figure}[!]
	\centering
	\includegraphics[width=0.5\textwidth]{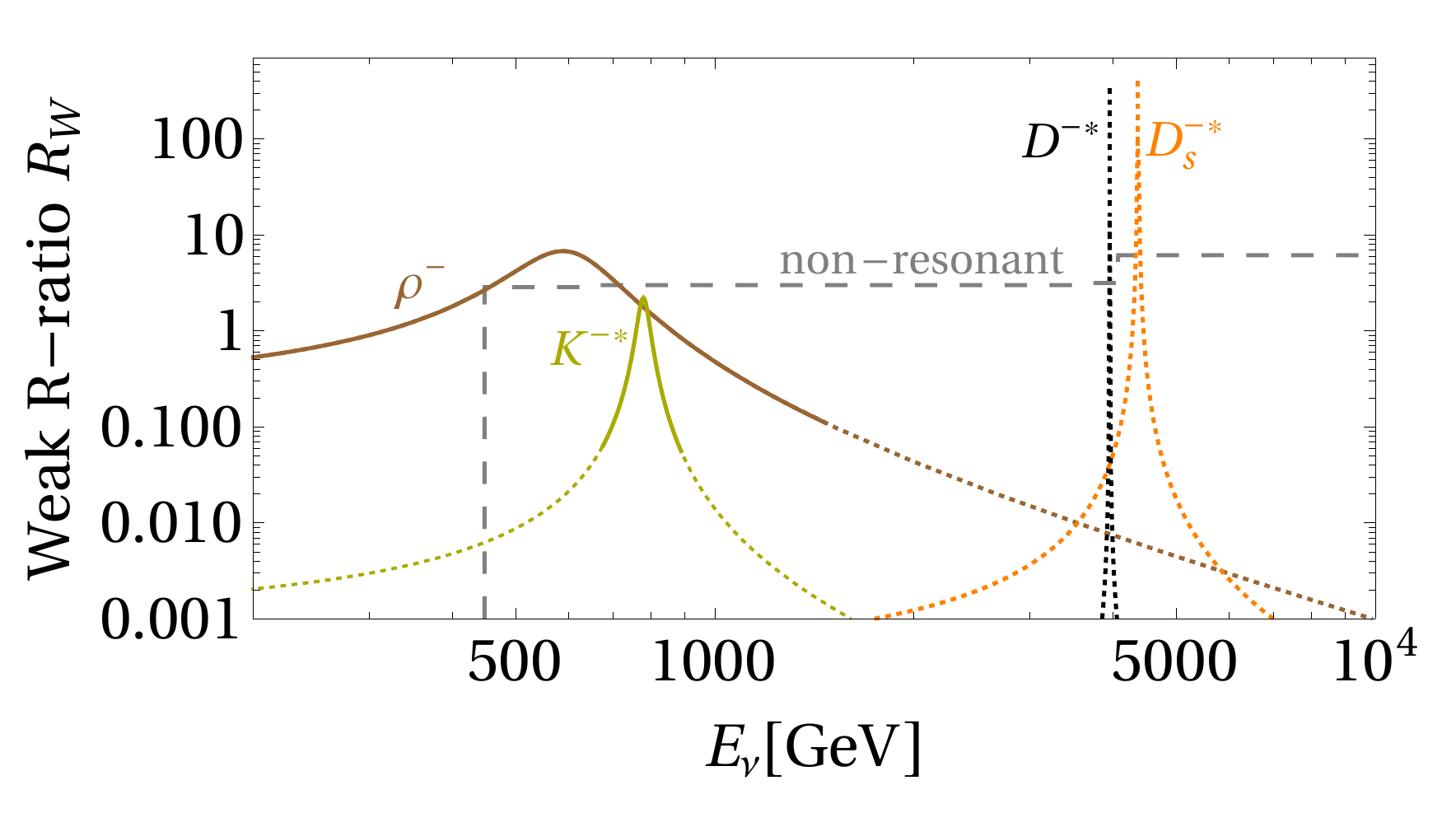} 
	\caption{Resonant contributions to the $R_W$ cross section ratio,  $\sigma(\bar{\nu}_e e^-\to \text{hadrons})/\sigma(\bar{\nu}_e e^-\to \bar{\nu}_\mu \mu^-)$,  for different charged mesons. The dashed line depicts the result of a naive perturbative parton-model calculation, with thresholds at the constituent quark masses. For the wider resonances $\rho^-$ and $K^{*-}$, the region between $E_\nu^\pm=(m\pm 3\Gamma)^2/(2m_e)$ is drawn with a solid line. 
%The chiral-suppressed pseudoscalar meson cross section is illustrated as well; shown $\pi^-$ cross section is multiplied by $10$ orders of magnitude.
}
	\label{fig:Rratio}
\end{figure}

The relevant contributions to $R_W$  are depicted in \cref{fig:Rratio}. For the denominator we take $\sigma(\bar{\nu}_\mu \mu^-)=(96 \pi)^{-1} g^4 \, s \, (M_W^2 - s)^{-2}$, where $s=2 E_\nu m_e$ and $g$ and $M_W$ are the weak coupling and the $W$-boson mass, respectively. To compute the numerator of $R_W$, we focus on resonant energies and use the Breit-Wigner formula~\cite{ParticleDataGroup:2020ssz} with ${\rm Br}_{\rm fi} =1$ for all mesons depicted in \cref{fig:Rratio}.
In addition to the mesonic resonances, we also include the naive constituent-quark-level QCD estimate  (dashed gray line), assuming constituent masses of 336~MeV, 340~MeV, and 486~MeV for the up, down, and strange quarks, respectively. 
 %%%%%%%%%%%%%%%%%%%%%%%%%%%%%%%%%%%%%%%%%%%%%%%%%%%%%%%%%%%%%%%%%%%%%%%%%%%%%%%%%%%%%%%%%%%%%%%%%
 %%%%%%%%%%%%%%%%%%%%%%%%%%%%%%%%%%%%%%%%%%%%%%%%%%%%%%%%%%%%%%%%%%%%%%%%%%%%%%%%%%%%%%%%%%%%%%%%

\textbf{Event rate estimates.} As alluded to in the introduction, there are two promising sources of neutrinos with $E_\nu \gg 100$ GeV that offer the opportunity to detect meson resonances. First, atmospheric and astrophysical neutrinos offer a broad, steeply-falling flux  with energies up to the PeV regime. As we argue below, these are sufficiently numerous to produce $O(1)$ events in IceCube per year.  Second, neutrinos produced at the LHC  offer a high intensity beam with energies ranging from $\sim { \rm 100~GeV -3~TeV}$ \cite{Anchordoqui:2021ghd}, these neutrinos can be detected using both FASER$\nu$ \cite{FASER:2020gpr} and SND@LHC \cite{SHIP:2020sos}; for our proposed signal SND@LHC does not yield observable event rates, and so we focus on FASER$\nu$ in what follows. We note that FASER$\nu$ has recently demonstrated the ability to detect and measure neutrino interactions \cite{FASER:2021mtu}. The planned FPF offers the advantage of high-resolution and larger detectors, including FASER$\nu$2's nuclear emulsion detector, and the potential opportunity to deploy a LArTPC further downstream (see \cite{Batell:2021blf} for the FLArE proposal).
In \cref{table}, we list the expected number of $\rho$ and $K^*$ resonant events in the aforementioned experiments. IceCube has a sub-detector volume with a higher density of digital optical modules, which allow for a lower energy threshold and better spatial resolution. This sub-volume is referred to as DeepCore \cite{IceCube:2011ucd} and we calculate event rates both for the full IceCube fiducial volume and for DeepCore with both estimates corresponding to 10 years of total exposure. We expect 10--100 $\rho^-$ and 1--10 $K^{-*}$ events in FASER$\nu$2 and in the two proposed liquid argon experimental setups, FLArE-10 and FLArE-100. For a comparison, the number of 
antineutrino induced  $D^{-*}$ and $D_s^{-*}$ resonances is $\sim 10^{-3}-10^{-2}$.
These detectors are envisioned to collect data during the High-Luminosity Large Hadron Collider (HL-LHC) stage \cite{Apollinari:2015wtw}; an integrated luminosity of $3000~ {\rm fb}^{-1}$ is expected for the period 2027--2038.

\begin{table}[t!]

{\color{black}{

{\small{
 \begin{tabular}{c|c|c|c|c}
\hline
\hspace{0.0125\linewidth} Experiment \hspace{0.0125\linewidth} &  $\rho^{-}$, $\pm\Gamma/2$ &  $\rho^{-}$, $\pm2\Gamma$ &  $K^{-*}$, $\pm\Gamma/2$ &  $K^{-*}$, $\pm2\Gamma$ \\ \hline 
FASER$\nu$ &  0.3 & 0.5  & -- & --   \\ 
FASER$\nu$2 & 23 & 37  &  0.7 & 3  \\ 
FLArE-10 &  11 & 19  &  0.3 & 2   \\
FLArE-100 & 63 & 103  &  2 & 8  \\ 
DeepCore & 3 (1) & 5 (2)  &  --  & --  \\ 
IceCube &  8 (40) & 17 (83)  &  -- & -- \\  \hline
\end{tabular}
}}
\caption{Estimated number of $\rho^-$ and $K^{-*}$ resonance-mediated events at different experimental setups.
% (for reference we expect fewer than $10^{-6}$ $\pi^-$ events at FLArE-100 and FASER$\nu$2)
    We show results for the cases where the integral in \cref{eq:rates} is within $\pm\Gamma/2$ and  $\pm2\Gamma$ of the resonance peak.
  A dash (--) indicates that less than $0.1$ events are expected. For IceCube and DeepCore, we use the effective masses given
  in Fig.~2 of \cite{ICmass} and consider 10 years of data taking. In parenthesis, we indicate the event rates at IceCube and DeepCore for the case when the
  effective mass matches the total mass. Event rates at present and proposed LHC-based detectors are computed using the fluxes from \cite{FASER:2019dxq} and the experimental configurations from
  \cite{Batell:2021blf}.}
\label{table}

}}
\end{table}

\textbf{Experimental signatures.} 
The estimates in \cref{table} do not include the impact of realistic background mitigation strategies. In what follows, we discuss some distinguishing characteristics of resonant meson production and the capabilities of nuclear emulsion, LArTPC, and Cherenkov detectors. 

We start with the basic kinematic properties of $\bar{\nu}_e e^-$ scattering and contrast them to DIS, which dominates neutrino--matter scattering for $E_{\nu}\sim 1$~TeV. At these neutrino energies, to a good approximation, nuclear structure can be neglected and one can concentrate on neutrino interactions with free nucleons $N$. The collision is most clearly described in the center-of-mass frame of the $\nu N$ system, which is connected to the lab frame by the Lorentz factor $\gamma_{\rm cm} = \sqrt{2 E_\nu/ m_N} \sim 36 \sqrt{E_\nu/600~{\rm GeV}}$. In the center-of-mass frame, both longitudinal and transverse momenta are proportional to $\sqrt{s}$ such that the typical scattering angle in the laboratory frame is $\theta_{\nu N} \sim 1/\gamma_{\rm cm} \sim {\rm 28~mrad} \times \sqrt{600~{\rm GeV/E_\nu}}$. When antineutrinos scatter on electrons, the boost between the center of mass and laboratory frame is, $\gamma_{\rm cm} \sim 1500 \sqrt{E_\nu/600~{\rm GeV}}$, larger by a factor of $\sqrt{m_N/m_e} \sim 43$, resulting in a typical angular scale of $\theta_{\nu e} \sim {\rm 0.7~mrad} \sqrt{600~{\rm GeV}/E_\nu}$. Therefore, detectors with mrad angular resolution can easily distinguish between these two different scenarios. 

The main resonance-mediated final state for a $\rho^-$ resonance (at $E_\nu \approx 580$ GeV) is $\pi^-\pi^0$, with branching ratio close to 100\%. 
Interestingly, the system is so boosted that the $\pi^0$ is, typically, not prompt: its decay length is $\lambda_{\pi^0} = \gamma\beta c\tau_{\pi^0} \sim (300 {\rm GeV}/0.135 ~{\rm GeV}) \times 25~{\rm nm} \sim 60 ~{\rm \mu m}$, leading to a visible displaced vertex in detectors with spatial resolution of $10~{\rm \mu m}$ or better. The subsequent $\pi^0$ decay will result in a $\gamma\gamma$ pair, each of which will convert to an $e^+e^-$-pair in a (typically macroscopic) length scale of order one radiation length $X_0$ of the detector material for $E_\gamma \sim 100~{\rm GeV}$. The conversion lengths of the two photons are realized in stochastic processes such that  one photon will convert before the other. Concurrently, the daughter $\pi^-$  will be ultra-relativistic ($\beta\gamma \gtrsim 1000$) such that radiative energy loss processes dominate over those induced by Bethe-Bloch ionization, and $\dd E/\dd x$ can be an order of magnitude larger than that of a minimally ionizing particle. 

\begin{figure}  
   \includegraphics[width=\linewidth]{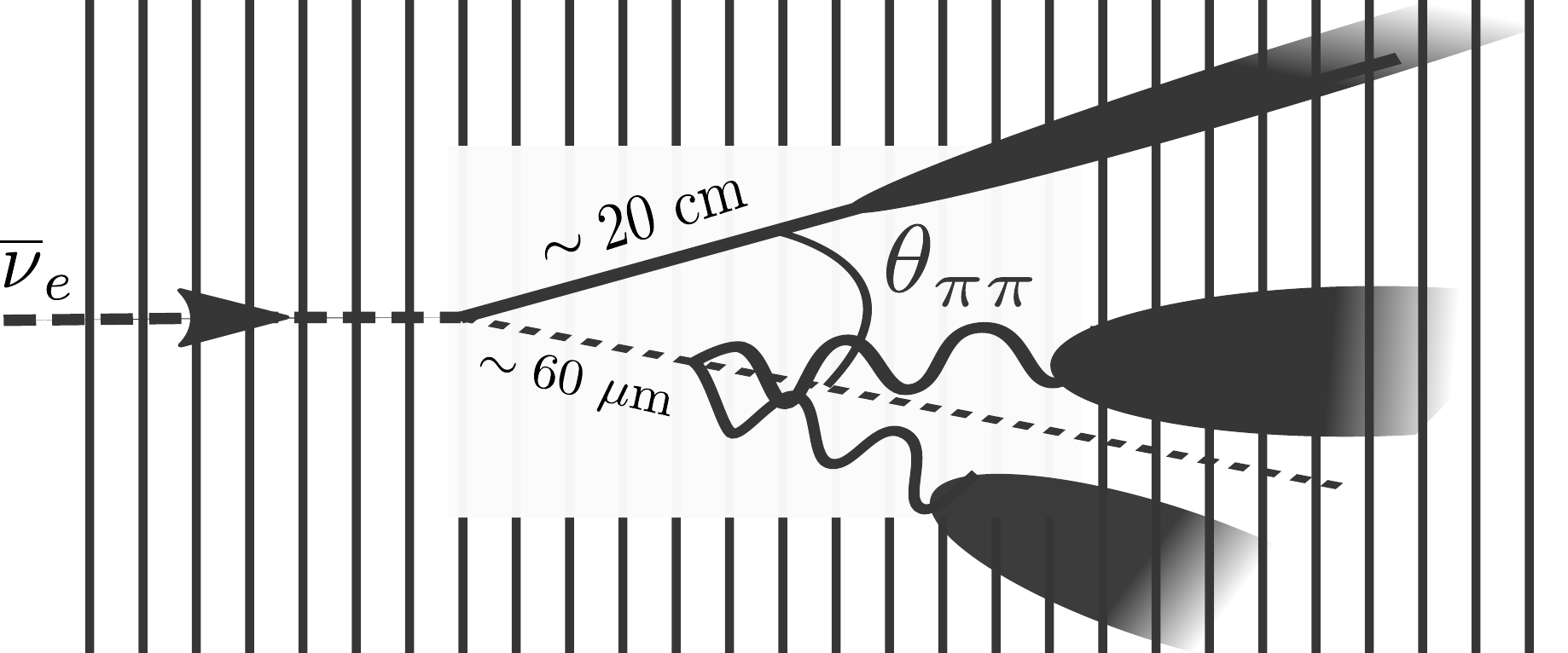}
    \caption{A cartoon of the typical event topology for $\bar{\nu}_e e^-\rightarrow \pi^-\pi^0$ in a nuclear emulsion detector. Two photons (i.e.\ EM showers) define a vertex displaced by, approximately, $60~\mu{\rm m}$ from a $\pi^-$ track, which has associated EM activity from hard bremsstrahlung. The showers are displaced by roughly one radiation length from the reconstructed vertex. The opening angle is $\theta_{\pi\pi}\sim m_e/m_\rho \sim 1/1500$. \label{cartoon} }
\end{figure}

FASER$\nu$ is a 1.2 tonne detector  located 480 m downstream from the ATLAS interaction point at the LHC that contains emulsion films and tungsten plates~\cite{FASER:2019dxq}. This kind of detector has a remarkable capability to reconstruct charged-tracks and other energy depositions, with spatial resolution as good as 50~nm. FASER$\nu$ projects a 400 nm spatial resolution which translates to an angular resolution of 0.06~mrad for track lengths of roughly 1~cm \cite{FASER:2019dxq}. This allows the detector to identify, e.g.\ tau leptons, $D^\pm$ mesons and $B^\pm$ mesons via the ``kinks'' resulting from their decay inside the detector. The resonant $\rho^-$ signature would consist of a forward track from the $\pi^-$ and two displaced electromagnetic (EM) showers, coming from the two photons from the $\pi^0$ decay, see \cref{cartoon}. The showers would define a vertex slightly displaced from the start of the $\pi^-$ track. This unique event topology, combined with the absence of other charged tracks, should serve as a powerful veto of DIS events, typically associated to intense hadronic activity. 

Furthermore, as discussed above, the typical angle between the $\pi^0$ and $\pi^-$ is of order $0.7~{\rm mrad}$ compared to $28~{\rm mrad}$ for typical DIS-produced tracks. One can also reconstruct the invariant mass of the $\pi^0 \pi^-$ pair and require it to lie within $\Gamma_\rho \sim 150$ MeV of $m_\rho\approx 770$ MeV. Assuming the $\pi^0$ and $\pi^-$ tracks can be reliably identified, their invariant mass is $m_{\pi\pi}^2= m_{\pi^0}^2 + m_{\pi^-}^2 +  E_{\pi^0} E_{\pi^-} \theta_{\pi \pi}^2$, where we are working in the small-angle, ultra-relativistic  approximation. $E_{\pi^0}$ is the reconstructed energy of the $\gamma\gamma$ EM shower, and $\theta_{\pi\pi}$ is the angle between the $\pi^-$ track and the center of the $\pi^0$ induced EM shower. Taking $\delta E/E \sim 5\%/(E/100~\rm{GeV})^{1/2}$ for both $E_{\pi^-}$ and $E_{\pi^0}$ \cite{Juget:2009zz,FASER:2019dxq} and $\delta \theta_{\pi\pi} \sim 10\% $ and assuming that the angular resolution dominates the error budget \footnote{While we assume that $\delta E/E \sim 5\%/(E/100~\rm{GeV})^{1/2}$ as presented in \cite{Juget:2009zz}, Ref. \cite{FASER:2019dxq} notes that a larger number of low-energy background electron tracks produced by high-energy muons may limit the energy resolution. We assume this is not case, however this deserves further investigation.} we estimate $\delta m_{\pi\pi}^2/m_{\pi\pi}^2 \sim 15\%\times(\delta\theta_{\pi\pi}/10\%)$, sufficient to identify the $\rho^-$ peak even allowing for lower angular resolutions. This measurement can be combined with calorimetric information, which should allow one to reconstruct $E_\nu = m_\rho^2/2m_e \approx 580$  GeV providing two independent criteria with which to identify likely $\rho^-$ candidates.

In addition to the large DIS background which is common across all three experiments, FASER$\nu$ and FASER$\nu$2 must also contend with a large background from through-going muons. The estimates presented in \cite{FASER:2018bac} suggest a muon background of 2.5 Hz/cm$^2$, which translates to $10^5-10^6$ muon tracks per cm$^2$ in a given detector exposure. Our candidate event involves one charged pion track and two displaced electromagnetic showers from $\pi^0 \rightarrow\gamma\gamma$. No dedicated studies have been performed that guarantee that this signal can be isolated from the sizeable background of electromagnetic showers produced  by muons. Nevertheless, future research may reveal new strategies that could enable the isolation of our signal topology from the background or, alternatively, the muon background could be mitigated by using a sweeper magnet as suggested in \cite{Anchordoqui:2021ghd}. In what follows we assume that the muon background can be controlled, and focus our attention on background events from neutral current DIS.

A $\rho^-$ candidate will have one $\pi^-$-like track and one $\pi^0$-like pair of EM showers, and can therefore be reduced to three kinematic variables. We take $\omega= E_{\pi^-}+E_{\pi^0}$, $\theta_{\pi\pi}$, and $m_{\pi\pi}^2$ as a convenient linearly independent basis. First, require  $\omega$ to lie close to $580$ GeV. The neutral current DIS cross section is around 100 times larger than the $\rho^-$ production cross section near its resonant peak, and all of $\bar{\nu}_e,~\bar{\nu}_\mu, ~\nu_e,$ and $\nu_\mu$ contribute to DIS scattering. Since $\nu_\mu$ and $\bar{\nu}_\mu$ fluxes are around 10 times larger than those of $\bar{\nu}_e$ and $\nu_e$, we anticipate a signal to background ratio of $\sim 1:5000$  imposing only a cut on $\omega$; this is an overestimate as it ignores particle identification (PID) of the exclusive final state $\pi^-\pi^0$. Next, require $\theta_{\pi\pi} \lesssim {\rm few ~mrad}$, as this  should provide an improvement in the signal to background ratio of order $\theta_{\nu N}/\theta_{\nu e} \sim 50$, with a negligible loss in signal events. This cut alone will improve the signal to background ratio to $\sim 1:100$ (again neglecting PID).  Further cuts may be imposed on charged-track multiplicity and photon multiplicity, as our signal has only one charged pion and two EM showers from the $\pi^0$, while typical neutral current DIS events will have much higher hadronic multiplicities. We anticipate a reduction in background by at least 50 (see Fig.\ 7 of \cite{Przybycien:1996zb}).
%, e.g.\ $n_\gamma <2$ 
Finally, define an additional cut on the reconstructed value of $m_{\pi\pi}$, as defined above, which should agree with the $\rho$ mass to within $\sim 20\%$ accuracy. Naively, this would suppress backgrounds by a further factor of 5, however $m_{\pi\pi}^2$ is correlated with $E_{\pi^+} + E_{\pi^-}$ and $\theta_{\pi\pi}$ and so there is some redundancy in this variable, and the background suppression from such a cut will likely be weaker. Nevertheless, provided this final cut can suppress backgrounds by at least a factor of 2, then a signal to background ratio comfortably larger than one is achievable, while simultaneously keeping most of the signal events.

The FLArE proposal consists of a LArTPC deployed 620~m from the ATLAS interaction point~\cite{Batell:2021blf}. While LArTPC detectors do not have the granularity of emulsion detectors, they still have 3~mm spatial resolution and their calorimetry allows for efficient PID using $\dd E/\dd x$. The radiation length in liquid argon is $X_0\approx 17$ cm \cite{LAr}, and the angular resolution for a 1~cm track is around $100~{\rm mrad}$ \footnote{The spatial resolution in a LArTPC is limited by the wire spacing, of order 1~mm. We estimate the angular resolution as $1~{\rm mm}/({\rm track~length})$.}. 
Consequently, the $\pi^-$ track and the $\gamma\gamma$ pair from  the $\pi^0$ decay generally overlap. Nevertheless, the track will have a distinct ionization profile. % $\dd E/\dd x$ along its longitudinal axis. 
In more detail, for the first $10~{\rm cm}$ or so only the $\pi^-$ will be visible. Its kinematics,  $\beta\gamma \sim 3000$, are such that bremsstrahlung will dominate over ionization but the emitted photons will not appear until after a full radiation length and only the ionization signature will be visible. Further down the track, each  $\gamma$ from the $\pi^0$ decay will convert to $e^+e^-$, each at a different point. In addition to the $\gamma \gamma$ pair from $\pi^0$ decay, other hard bremsstrahlung photons produced by the $\pi^-$ will also pair produce. The result will be a forward-pointing ``hadronic flashlight'' that has discrete jumps in $\dd E/\dd x$ as a function of $x$ with no associated wide-angle hadronic activity from the interaction vertex. 

Cherenkov detectors have poorer angular resolution relative to LArTPCs. Therefore, the $\pi^-$, and the two $\gamma$s from the $\pi^0$ decay will result in a single collimated event. In water, for the energies of interest, the typical hadronic interaction length is $\lambda \sim 90$~cm and a radiation length is $X_0\sim 30$~cm. Both hadronic and EM cascades are on the order of $5-10$ interaction lengths and therefore sufficiently short that the $\rho^-$-mediated events considered here would be classified as ``fully contained cascades'' in the IceCube nomenclature. This class of events has order $10\%$ energy resolution so the $\rho^-$ resonance region, $540~{\rm GeV} \lesssim E_\nu \lesssim 640$~GeV, can be realistically isolated.  

Our estimates in \cref{table} suggest that there are around ten resonant $\pi^-\pi^0$ events in IceCube's existing event catalogue so it is worthwhile to understand if these events can be reliably identified. Typical cascade events involve $\nu_e$ or $\bar{\nu}_e$ charged-current scattering or neutral-current events. In both cases the outgoing lepton (electron or neutrino) typically carries away around $80\%$ of the energy while the remaining $20\%$ is transferred to the hadronic system. Concentrating on events where the reconstructed neutrino energy lies inside the $\rho^-$ resonance window, a neutral-current event would need to deposit at least $W_0\sim 540$ GeV of hadronic energy, corresponding to a typical neutrino energy $E_\nu\sim 2.7$ TeV. Charged-current backgrounds, instead, involve neutrinos with $E_{\nu}\sim 600~{\rm GeV}$, since both the electron/positron are also fully contained in the cascade. Using the atmospheric neutrino fluxes from \cite{Hondaflux} and accounting for the energy dependency of the cross sections, we estimate a signal to background ratio of order 1:40 \footnote{The $\bar{\nu}_e$ differential flux at $580$ GeV, responsible for $\rho^-$ resonant production, is $\dd \Phi_{\bar{\nu}_e}/\dd E_\nu \sim 5\times 10^{-8} ~{\rm m}^{-2} {\rm s}^{-1} {\rm sr}^{-1} {\rm GeV}^{-1}$. The atmospheric neutrino differential flux (all flavors) at  $2.7$ TeV is roughly $\dd \Phi_\nu/\dd E_\nu \sim 1\times 10^{-8} ~{\rm m}^{-2} {\rm s}^{-1} {\rm sr}^{-1} {\rm GeV}^{-1}$ \cite{Hondaflux}, approximately five times smaller than the relevant $\bar{\nu}_e$ flux. The  neutral-current DIS cross section is given roughly by $\sigma_{\rm NC}({\rm DIS})\sim (E_\nu/2~\rm{TeV})\times 10^{-35} {\rm cm}^2$ per nucleon \cite{Ismail:2020yqc}, roughly two orders of magnitude larger than that of $\rho^-$ resonant production, $\sigma_{\rm res}(\rho^-) \sim 10^{-37}~{\rm cm}^2$. Taking into account that there are roughly twice as many nucleons as electrons in water, this yields $S:B\sim 1:40$.} for neutral-current backgrounds and 1:140 \footnote{Accounting for the slightly larger flux of $\nu_e$ relative to $\bar{\nu}_e$ and the fact that the cross section is twice as large, we find an effective cross section of $\sigma_{\rm eff} \approx  3.5 \times \sigma_{\rm DIS}(\bar{\nu}_e) = 7\times 10^{-36}~{\rm cm}^2$ per nucleon, roughly $70$ times larger than the resonant production cross section. Taking into account that there are roughly twice as many nucleons as electrons in water, this yields $S:B\sim 1:140$.} for charged current backgrounds.

\textbf{Conclusions.} The production of charged-meson resonances in $\bar{\nu}_e-e$ scattering is an interesting and previously inaccessible SM neutrino reaction. Existing data from IceCube may already hold a handful of $\rho^-$ resonances, but these lie beneath a sizable DIS background. In contrast, the situation at the LHC's FPF appears to be more promising. We estimate 10--100 total meson resonance events at proposed FPF detectors and, while backgrounds are naively also large, their excellent spatial and angular resolutions, in the case of nuclear emulsions, or calorimetry, in the case of LArTPCs, may allow for very effective background rejection while maintaining order one signal efficiency. It may be interesting to consider the optimal rapidity of an emulsion detector for the purpose of detecting a $\rho^-$ resonance. In \cite{Kling:2021gos} the authors note that the spectrum of neutrinos varies with rapidity, and it is conceivable that there exists a detector placement which would supply a flux of neutrinos that peaks at the energy necessary to produce an on-shell $\rho^-$, however this lies beyond the scope of the current work. Meson resonances offer an interesting case study that illustrates the capabilities of both nuclear emulsion detectors and LArTPCs and can serve as an intriguing physics target for the LHC's FPF.

\textbf{Acknowledgements.}
We would like to thank Bhupal Dev, Felix Kling, Jonathan Rosner and Sebastian Trojanowski for very useful discussions, and to the anonymous referee for helpful feedback and suggestions. RP acknowldges the hospitality of the Fermilab theory group. This work was supported in part by the US Department of Energy (DOE) through grants \#de-sc0010143 and \#de-se0019095, and in part by the NSF grant PHY-1630782. The document was prepared using the resources of the Fermi National Accelerator Laboratory (Fermilab), a DOE, Office of Science, HEP User Facility. Fermilab is managed by Fermi Research Alliance, LLC (FRA), acting under Contract No.\ DE-AC02-07CH11359.
%%%%%%%%%%%%%

%\bibliographystyle{JHEP}
\bibliography{glashow}

\end{document}